\documentclass[aps,prl,preprint,showpacs]{revtex4}
\usepackage{mathrsfs}
\usepackage{graphicx}
\usepackage{amssymb}
\usepackage{amsmath}

\usepackage{setspace}

\begin{document}

%%%%%%%%%%%%%%%%%% title page information %%%%%%%%%%%%%%%%%%
\title{Near-infrared single-photons from aligned molecules in ultrathin crystalline films at room temperature}

\author{C. Toninelli$^{1}$, K. Early$^{1, 2}$, J. Bremi$^{1}$, A. Renn$^{1}$, \\
S. G\"{o}tzinger$^{1}$, and V. Sandoghdar$^{1}$}

\affiliation{$^{1}$Laboratory of Physical Chemistry and optETH, ETH
         Z\"{u}rich, CH-8093 Z\"{u}rich, Switzerland   \\
         $^{2}$The George Richardson, Jr.\,Chemistry Research Laboratory,
         Department of Chemistry, University of Massachusetts, Amherst, Massachusetts 01003, USA}

\begin{abstract}
We investigate the optical properties of Dibenzoterrylene (DBT)
molecules in a spin-coated crystalline film of anthracence. By
performing single molecule studies, we show that the dipole moments
of the DBT  molecules are oriented parallel to the plane of the
film. Despite a film thickness of only $20$\,nm, we observe an
exceptional photostability at room temperature and photon count
rates around $10^{^6}$ per second from a single molecule. These
properties together with an emission wavelength around $800$\,nm
make this system attractive for applications in nanophotonics and
quantum optics.
\end{abstract}

\maketitle

Room temperature single-photon sources are desirable for a variety
of applications, ranging from quantum key distribution to building a
standard to measure the luminous intensity of a light source
\cite{Lounis2005,Scheel2009}. Solid state systems are especially
appealing, not only because they are easy-to-use but also because of
their potential for integration and scalability. However, most solid
state emitters suffer from a limited photostability. So far only
nitrogen-vacancy centers in diamond \cite{kurtsiefer2000} and
terrylene molecules in a para-terphenyl host
\cite{Fleury1998,Lounis2000} have shown at room temperature stable
single-photon emission over extended periods of time.

Several years ago we discovered that terrylene is also extremely
photostable in a crystalline p-terphenyl film as thin as $20$
molecular layers \cite{Pfab2004}, which protects the molecule
against quencher agents such as oxygen \cite{Renn2006}. Doping
single emitters into thin films has several advantages, especially
in the context of single emitter experiments. First, background
fluorescence is strongly suppressed due to the optimized ratio
between emitter and matrix. Second, thin films are inherently
compatible with nanostructures such as plasmonic waveguides, since
the emitter is by definition in the near-field. Furthermore they can
easily be integrated into microcavities, either as part of a layered
structure in a linear cavity or via evanescent coupling to a
whispering-gallery resonator or a photonic crystal defect cavity.

\begin{figure}
\begin{center}
\includegraphics[width=0.8\textwidth]{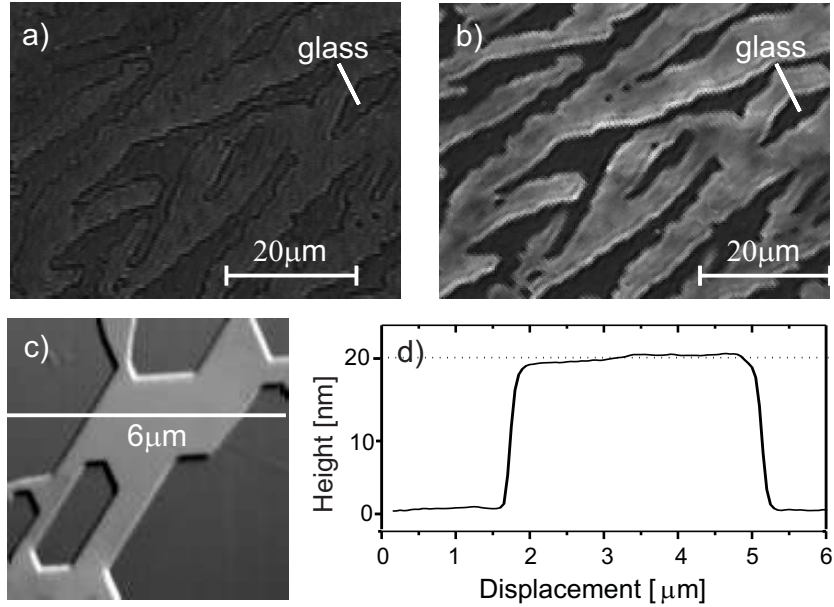}
\caption{(a,b) Polarization microscope images of a thin AC film,
spin coated on a glass cover slip. The analyzer is oriented
perpendicular to the polarizer. In (a) one of the main crystal axes
is aligned with the incoming polarized light. The contrast between
cover glass and crystalline features is therefore low. In (b) the
sample is rotated by $45^{\circ}$. This leads to a birefringence of
the crystalline AC film which rotates the polarization of the
incoming light. The contrast is therefore maximal. (c) AFM
topography image of the sample. The well defined growth angles give
further evidence for the crystalline nature of the film. (d) Cross
section as indicated in (c). The sample is typically flat with a
height of a few tens of nanometers.}\label{Fig1}
\end{center}
\end{figure}

In this paper we investigate Dibenzoterrylene (DBT) molecules, which
have been previously studied in crystals at cryogenic temperatures
\cite{Jelezko1996,Hofmann2005,Nicolet2007,Trebbia2009}. Here we
report on the fabrication of ultrathin crystalline anthracene (AC)
by a simple spin coating procedure on glass cover slides. To produce
the desired films, we prepared a solution of AC in diethyl ether
with a concentration of $2.5$\,mg/ml and added $10 $\,$\mu$l/ml of
benzene. The latter serves to improve the quality of the crystals
obtained from the spin-coating process. DBT was then dissolved in
toluene to obtain a $10$\,$\mu$M solution, which was further diluted
by a factor of $100$ with the AC/diethyl ether mixture. Then we spin
casted $20$\,$\mu$l of the solution containing AC and DBT onto a
glass cover slide. With a two-step process ($30$ s at $3000$\,RPM
followed by $20$\,s at $1500$\,RPM) on a commercial spin coater  we
obtained areas with crystalline islands that covered several
mm$^{2}$ of the substrate. Figure \ref{Fig1}\,(a) shows an optical
polarization microscope image of a typical sample area, containing
both film and bare glass regions. The contrast between glass and
crystalline film is very low since one of the main axes of the
anthracene crystal is aligned with the polarization vector of the
incoming light. In Fig. \ref{Fig1}\,(b) the same portion of the
crystal is rotated by $45^{\circ}$. In contrast to the amorphous
glass the crystal shows some birefringence. As a result, the
polarization vector of the transmitted light is rotated and the
contrast to glass is increased. We thus conclude that the AC film is
crystalline with the same optical axis over hundreds of square
microns. To obtain information about the topography of the host
matrix we performed atomic force microscopy (AFM). A typical
measurement is displayed in Fig. \ref{Fig1}\,(c), where well defined
crystalline structures are visible. In Fig. \ref{Fig1}\,(d) a cross
section is plotted, showing a fairly constant thickness of about
$20$\,nm.

\begin{figure}
\begin{center}
\includegraphics[width=\textwidth]{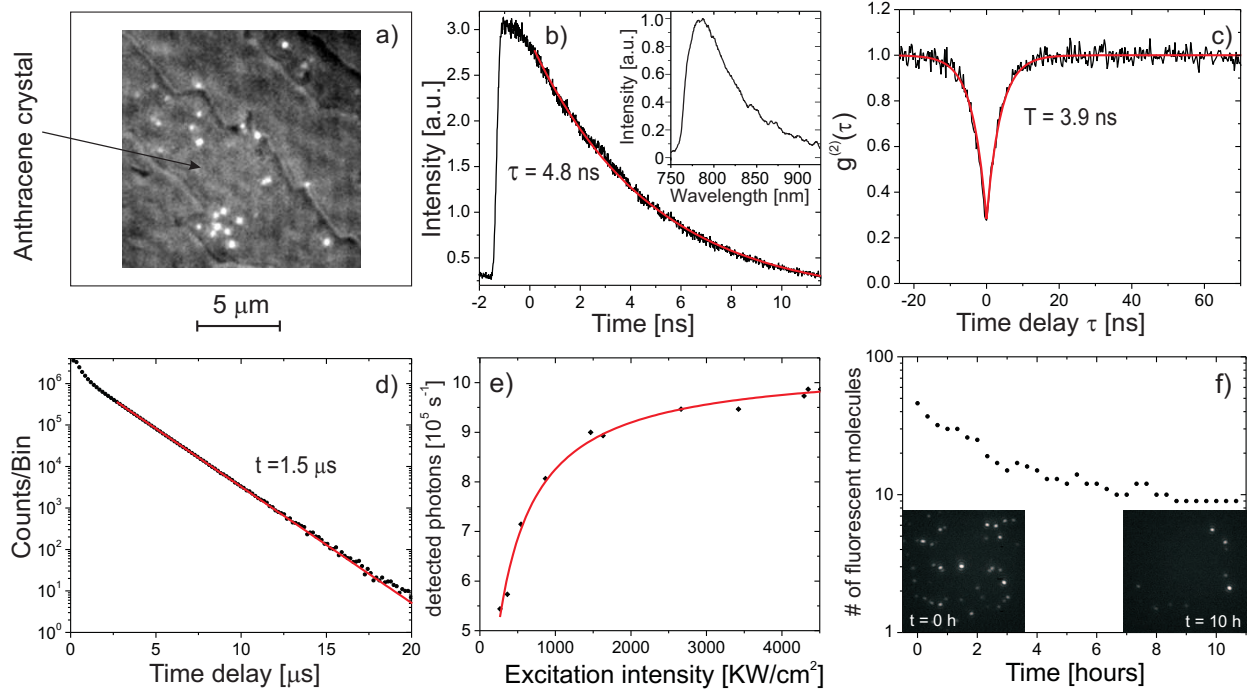}
\caption{ (a) Wide field image, where the sample was simultaneously
illuminated by a white-light source and a laser. Individual
molecules are clearly visible within the crystalline domains. (b)
Fluorescence lifetime measurement on a DBT molecule. The red curve
represents an exponential fit to the experimental data, yielding a
decay time of $4.8$\,ns. Inset: fluorescence spectrum of a single
DBT molecule. (c) Photon-correlation measurement under CW
excitation. Strong anti-bunching is observable. The red curve is a
fit to the experimental data. (d) Histogram of the inter-photon
arrival times. The obtained decay time yields a $1.5$\,$\mu$s
lifetime of the triplet state. (e) Saturation measurement: number of
detected photons per second depending on the pump power. The red
curve is a two-level model fit to the saturation behavior
\cite{Lounis2000}. (f) Photostability of DBT molecules: the insets
show wide field images at the beginning of the measurement and after
$10$\,hours of continuous illumination. Ten molecules out of $43$
could not be photobleached.}\label{Fig2}\end{center}
\end{figure}

The optical investigations of DBT were carried out on single
molecules in thin films by means of fluorescence microscopy. Our
fluorescence microscopy setup was equipped with a continuous wave
(CW) and a pulsed Ti:Sapphire laser ($120$\,fs pulse width) to
efficiently excite the molecules at a wavelength of $725$\,nm using
an oil immersion objective (N.A. $1.4$). A lens could be inserted in
the excitation path to switch between confocal and wide-field
illumination. Fluorescence was then collected by the same objective
and separated from the excitation light with a longpass filter.
Several detection paths allowed access to a CCD camera, a
fiber-coupled avalanche photodiode (APD), a spectrometer or a
Hanbury-Brown-Twiss (HBT) photon correlator.

Figure \ref{Fig2}\,(a) shows a wide-field CCD camera image of DBT
molecules in an AC film. Individual molecules can be clearly
distinguished. By switching to confocal excitation, we selected
individual molecules for further investigations. The inset
 in Fig, \ref{Fig2}\,(b) displays the fluorescence spectrum of a DBT
molecule. It has its maximum at $790$\,nm and a width of about
$50$\,nm. However, because the detection efficiency of the
spectrometer drops between $850$\,nm and $900$\,nm by more than a
factor of two, the spectrum is slightly distorted in this wavelength
range. To gain further information on the molecule's properties, we
directed the photons generated by pulsed excitation on an APD and
applied a time correlated single-photon counting technique to
determine the lifetime of the excited state. Figure \ref{Fig2}\,(b)
shows an example of a time-resolved intensity measurement, which
could be fitted with a single exponential decay. Repeating the
measurement on several molecules, we obtained lifetimes between
$3.3$\,ns and $5.7$\,ns. Then we employed photon correlation
measurements using the HBT setup to verify the identification of
isolated single DBT molecules. The CW photon autocorrelation
measurement in Fig. \ref{Fig2}\,(c) shows a dip in the second order
correlation function at delay $\tau=0$, which corresponds to a
reduction of the two-photon detection probability to $0.28$. This
result motivates the use of DBT as a near-infrared single-photon
source.

The performance of a single-photon source is in many cases
compromised by fluorescence intermittency. The blinking of
semiconductor nanocrystals is a well-known example of this
phenomenon \cite{Verberk2002}. In the case of molecules one has to
worry about a long-lived triplet state, populated by intersystem
crossing, which can interrupt the continuous stream of photons. To
determine the lifetime of the triplet state, we recorded a histogram
of the inter-photon arrival times with a pump rate higher than the
triplet decay rate \cite{Bernard1993}. Under this condition the dark
intervals in the fluorescence are limited by the triplet lifetime,
which can then be extracted from the slow decay in \ref{Fig2}\,(d).
The initial fast decay is a measure for the pump rate. Considering
the obtained triplet lifetime of $1.5$\,$\mu$s together with an
extremely low intersystem crossing yield of $10^{-7}$
\cite{Nicolet2007}, we can neglect the effect of the triplet state
on the efficiency of a DBT single-photon source. Another important
property is the brightness, which can be extracted from saturation
measurements. Fig.\ref{Fig2}\,(e) shows that we can detect almost to
one million photons per second at pump intensities close to
saturation. Such count rates are among the highest ever reported
\cite{Strauf2007}. By considering the maximum count rate together
with the determined lifetime, we can deduce a total detection
efficiency of $0.5\,\%$.

The photostability of DBT molecules embedded in thin crystalline AC
films is especially noteworthy, when considering that other
molecular emitters typically photobleach after $10^{4}-10^{7}$
photon emissions \cite{Eggeling1998}. We investigated the stability
of DBT by irradiating a sample continuously with an intensity of
$30$\,kW/cm$^2$ and recorded a wide-field image every $20$\,min over
a time period of more than $10$\,hours (see Fig. \ref{Fig2}\,(f)).
For about $30$ out of $40$ molecules we could attribute a
'half-life' of $4$\,h by fitting an exponential decay to the
experimental data. These molecules emitted more than $10^{12}$
photons before photobleaching, assuming the above-calculated
detection efficiency of $0.5\%$. The remaining ten molecules,
however, did not suffer from any photobeaching, even after more than
$10$\,hours of constant illumination.

\begin{figure}[t!]
\begin{center}
\includegraphics[width= \textwidth]{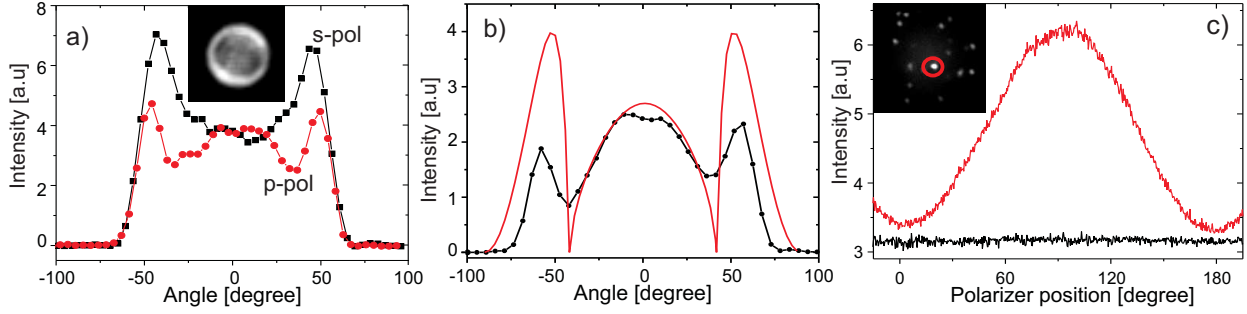}
\caption{(a) Inset: Back focal plane image of a single molecule.
Angular distribution of the emitted photons for two cross sections
which correspond to s and p polarization. (b) Emission pattern of
p-polarized light from an ensemble of molecules, fitted with the
angular distribution of a single dipole one degree out of plane (red
curve). (c) Dependency of the detected fluorescence intensity of a
single molecule on the orientation of a polarizer in the detection
path.}\label{Fig3}
\end{center}
\end{figure}

A further feature of our sample is the fixed orientation of the
molecular dipole moment parallel to the cover glass. To investigate
the dipole orientation we utilized a back focal plane imaging
technique, which allows us to study the angular emission pattern of
single molecules \cite{Lieb2004}. Cross sections through the back
focal plane of an exemplary image along the two orthogonal
polarization axes are plotted in Fig. \ref{Fig3}\,(a). The symmetry
of the obtained shape indicates a horizontally aligned molecule. We
found that all molecules within an area of a few tens of microns
showed similar emission patterns and were out of plane at most by a
few degrees. The symbols in Fig. \ref{Fig3}\,(b) show that the sum
of the signals from many molecules was also centered around zero.
The solid curve in this figure displays the theoretically expected
pattern for a single dipole at the interface. The central part of
the angular pattern shows a very good agreement with the
experimental results and was fitted. The experimental side lobes
miss the fast modulations due to the finite angular resolution of
the experiment. They also fall short of the theoretical prediction
because the latter did not take into account the exact distance of
the molecule from the AC-air interface, which sensitively determines
how much light is emitted at angles beyond the critical angle. As a
second check for the alignment of the molecules, we performed
measurements where the orientation of a polarizer in the detection
path was varied. Figure \ref{Fig3}\,(c) shows that the fluorescence
signal of a single DBT molecule could be varied with a visibility of
$97$\,\%. We note in passing that the maximum detected fluorescence
occurred at similar polarizer positions for molecules in the same
field of view, supporting the fact that large crystalline domains
exist.

In conclusion, we have prepared by a simple spin coating procedure
ultrathin crystalline AC films doped with DBT. An analysis of single
molecule fluorescence reveals that DBT is horizontally aligned,
exceptionally photostable and bright. The near-infrared emission
wavelength of $800$\,nm is in many cases advantageous. Microcavities
are easier to fabricate for longer operation wavelengths and the
losses in gold or silver plasmonic structures are significantly
reduced. Furthermore, the orientation of the molecules can be
exploited to efficiently couple the emitted photons to any of the
above mentioned photonic structures, which makes this molecule
extremely attractive as easy-to-use active emitter in nanophotonics
and quantum optics.

This work was supported by the ETH Zurich via the INIT program
Quantum Systems for Information Technology (QSIT) and the Swiss
National Science Foundation. K.E. acknowledges support from the NSF
IGERT Program (DGE-$0504485$).

%%%%%%%%%%%%%%%%%%%%%%% References %%%%%%%%%%%%%%%%%%%%%%%%%
%\begin{thebibliography}{99}
%\bibliographystyle{osajnl}

\end{document}